\documentclass{ws-procs9x6-cpt16}

\begin{document}

\newcommand{\refeq}[1]{(\ref{#1})}
\def\etal {{\it et al.}}

\title{Improved Tests of Lorentz Invariance in the Matter Sector using\\
Atomic Clocks}

\author{H.\ Pihan-Le Bars,$^1$ C.\ Guerlin,$^{1,2}$ Q.G.\ Bailey,$^3$ 
S.\ Bize,$^1$ and P.\ Wolf$^1$}

\address{$^1$SYRTE, Observatoire de Paris, PSL Research University, CNRS\\
Sorbonne Universit\'es, UPMC Univ.\ Paris 06, LNE\\
61 avenue de l'Observatoire, 75014 Paris, France}

\address{$^2$Laboratoire Kastler Brossel, ENS-PSL Research University, CNRS \\ 
UPMC-Sorbonne Universit\'es, Coll\`ege de France,
75005 Paris, France}

\address{$^3$Embry-Riddle Aeronautical University,
Prescott, Arizona 86301, USA}

\begin{abstract}
For the purpose of searching for Lorentz-invariance violation in the minimal Standard-Model Extension, we perfom a reanalysis 
of data obtained from the $^{133}$Cs fountain clock
operating at SYRTE.
The previous study led to new limits on eight components 
of the $\tilde{c}_{\mu \nu}$ tensor, 
which quantifies the anisotropy of the proton's kinetic energy. 
We recently derived an advanced model for the frequency shift 
of hyperfine Zeeman transition due to Lorentz violation 
and became able to constrain the ninth component, 
the isotropic coefficient $\tilde{c}_{\text{{\tiny  \textsc{TT}}}}$, 
which is the least well-constrained coefficient of $\tilde{c}_{\mu \nu}$. 
This model is based on a second-order boost Lorentz transformation
from the laboratory frame to the Sun-centered frame,
and it gives rise to an improvement of five orders of magnitude 
on $\tilde{c}_{\text{{\tiny  \textsc{TT}}}}$ compared to the state of the art.
\end{abstract}

\bodymatter


\begin{figure}
\begin{center}
\includegraphics[width=0.6\hsize]{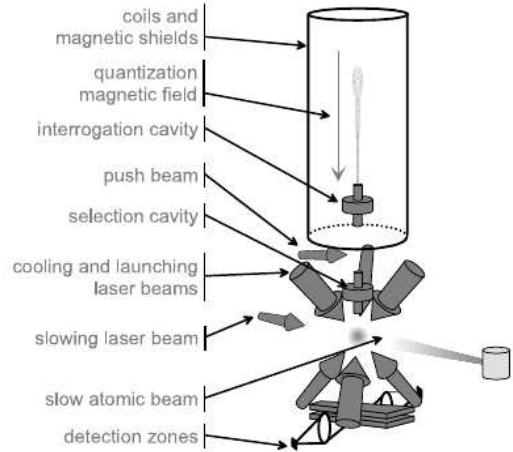}
\end{center}
\caption{Schematic view of an atomic fountain.\cite{Bize2005}}
\end{figure}


\phantom{}\vskip10pt\noindent
The $^{133}$Cs and $^{87}$Rb double fountain (see Fig.\ 1\cite{Bize2005}) 
was run in Cs mode on a combination of {$\vert F=3, m_F \rangle \longleftrightarrow \vert F=4, m_F\rangle$
hyperfine transitions,\cite{Guena2012,Guena2014}
which have good sensitivity to the quadrupolar energy shift of the proton and a weak dependence on the first-order Zeeman effect. The combined observable $\nu_c$, build by measuring quasi-simultaneously the clock frequency for $m_F = + 3, -3, 0$}, can be related to a model for hyperfine transitions 
in the minimal Standard-Model Extension (SME)\cite{Bluhm2003,Kostelecky1999} 
and leads to the laboratory-frame SME model 
presented in Ref.\ \refcite{Wolf2006}. 
This observable depends on the proton's laboratory-frame coefficient $\tilde{c}_q^p$, which is a combination of the $c_{\mu \nu}$ tensor components.


To search for a periodic modulation of the clock frequency, the laboratory coefficients must be expressed as functions of the Sun-centered frame coefficients.\cite{Kostelecky2002} This transformation is usually done via a first-order ($O(\beta)$) boost Lorentz transformation,\cite{Bluhm2003, Kostelecky1999, Wolf2006} but for purpose of setting a limit on the isotropic coefficient,  $\tilde{c}_{\text{{\tiny  \textsc{TT}}}}$, which appears in an {$O( \beta^2 )$} model suppressed by a factor $\beta^2$, we develop an improved model using a second-order boost matrix (see also Ref.\ \refcite{Guerlinproc}). This contains all the terms up to {$O( \beta^2 )$}, in contrast to Ref.\ \refcite{Hohensee2013} which kept the {$O( \beta^2 )$} terms exclusively for $\tilde{c}_{\text{{\tiny  \textsc{TT}}}}$. We also include the annual frequency, previously taken as a constant\cite{Wolf2006}. The model
now exhibits in total 13 frequency components (25 quadratures), instead of
3 frequency components (5 quadratures) for the previous analysis.


We perform a complete least-squares adjustment of the {$O( \beta^2 )$} model to the data used in Ref.\ \refcite{Wolf2006}. This model is fitted in the SME coefficient basis, which enables us to evaluate simultaneously the nine $\tilde{c}_{\mu \nu}$ coefficients for the proton and their respective correlations. It also avoids additional assumptions on parameter expectation values and underestimation of the uncertainties.\cite{HPB}
The main systematic effects are related to the first- and second-order Zeeman effects. The second-order effect is responsible for an offset of the data from zero, assessed at $-2.2$ mHz, and the residual first-order Zeeman effect is calibrated via a least-squares fitting of the {$O( \beta^2 )$} model to the time of flight of the atoms in the fountain.\cite{HPB,Wolf2006}

\begin{table}
\centering
{\footnotesize Table 1. \ Limits on SME Lorentz-violation coefficients 
$\tilde{c}$ for the proton in GeV.} 
\vskip 5pt
\begin{footnotesize}
$\begin{array}{c c c c c c c}
\hline \hline
\text{Coefficient}  &  \text{ Measured value }  & \multicolumn{3}{c}{\text{ Uncertainty }}   & \text{Unit (GeV)}\\
& &\text{ Statistical} & \text{Systematic} &\text{ Total }&\\
 \hline\vspace*{-3.2mm}\\ 
\tilde{c}_{\text{{\tiny  \textsc{Q}}}}& -0.3 	& 	10^{-2}	& 	2.1  & 	2.1	 & 10^{-22}  \vspace*{-.6mm}\\
\tilde{c}_{\text{{\tiny  \textsc{-}}}}&   1.4	& 	0.7	& 	8.9  & 	9.0	 & 10^{-24}  \vspace*{-.6mm}\\
\tilde{c}_{\text{{\tiny  \textsc{X}}}}&   -1.5 	& 	0.7	& 	5.2  & 5.3		 & 10^{-24}  \vspace*{-.6mm}\\
\tilde{c}_{\text{{\tiny  \textsc{Y}}}}&  	0.8  	&  	0.3	& 	1.6  & 	1.6	 & 10^{-24}  \vspace*{-.6mm}\\
\tilde{c}_{\text{{\tiny  \textsc{Z}}}}&  	1.0  	&  	0.8	& 	3.9  & 	3.9	 & 10^{-24}  \vspace*{-.6mm}\\
\tilde{c}_{\text{{\tiny  \textsc{TX}}}}&  -1.5  	& 	0.6	& 	5.7  & 	5.7	 & 10^{-20}  \vspace*{-.6mm}\\
\tilde{c}_{\text{{\tiny  \textsc{TY}}}}&   1.4	&	0.3	& 	5.9  & 5.9	 & 10^{-20}  \vspace*{-.6mm}\\
\tilde{c}_{\text{{\tiny  \textsc{TZ}}}}&   -1.1	&  	0.2	& 	3.5  & 3.5	 & 10^{-20}  \vspace*{-.6mm}\\
\tilde{c}_{\text{{\tiny  \textsc{TT}}}}&  1.6 	& 	0.9	& 	6.9  & 	6.9	 & 10^{-16}  \vspace*{-.4mm}\\
\hline
\hline
\end{array}$
\end{footnotesize}
\label{coeff}
\end{table}

The bounds on $\tilde{c}_{\mu \nu}$ components obtained using the complete {$O( \beta^2 )$} model are presented in Table 1. They show an improvement by five orders of magnitude on $\tilde{c}_{\text{{\tiny  \textsc{TT}}}}$ compared to the state of the art.\cite{datatables}
Despite our advanced model, the correlation matrix
still contains large values (up to $0.95$), except for the $\tilde{c}_{\text{{\tiny  \textsc{Q}}}}$ coefficient, which is almost decorrelated at this sensitivity level. This indicates that our marginalized uncertainties in Table 1 are dominated by those correlations, and could thus be significantly improved with more data spread over the year.



In conclusion, our improved model including $O\left( \beta^2 \right)$ terms and annual frequency modulations enables us to improve the present limits on the isotropic
coefficient $\tilde{c}_{\text{{\tiny  \textsc{TT}}}}$ by 5 orders of magnitude. Furthermore, we expect that an
additional data set would reduce the marginalized uncertainties and lead to
an improvement by one extra order of magitude on all the limits, bringing
the constraint on $\tilde{c}_{\text{{\tiny  \textsc{TT}}}}$ near one Planck scale suppresion, i.e. $10^{-17}$ GeV.

\end{document}